\documentclass[10pt,conference]{IEEEtran}
\usepackage{cite}
\usepackage{amsmath,amsfonts,amsthm,amssymb,amscd}
\usepackage{balance}
\usepackage{mathtools}
\usepackage{algorithm}
\usepackage{algpseudocode}
\usepackage{graphicx}
\usepackage{textcomp}
\usepackage{xcolor}
\usepackage{caption}
\usepackage{subcaption}
\usepackage{siunitx}
\usepackage{aligned-overset}
\usepackage{bbm}

\usepackage{breqn}

\def\BibTeX{{\rm B\kern-.05em{\sc i\kern-.025em b}\kern-.08em
    T\kern-.1667em\lower.7ex\hbox{E}\kern-.125emX}}

\newcommand{\mac}[1]{\underline{#1}}
\newcommand{\up}[1]{\overline{#1}}
\newcommand{\mbps}[1]{\SI[per-mode=fraction]{#1}{\mega\bit\per\second}}

\newcommand{\mb}[1]{\SI{#1}{\mega\bit}}

\newcommand{\ms}[1]{\SI{#1}{\milli\second}}

\theoremstyle{plain}
\newtheorem{thm}{Theorem}
\newtheorem{prop}[thm]{Proposition}

\theoremstyle{definition}
\newtheorem{defn}[thm]{Definition}
\theoremstyle{remark}
\newtheorem{rem}[thm]{Remark}
\theoremstyle{plain}
\newtheorem{lem}[thm]{\protect\lemmaname}

\newcommand{\conv}[3]{#1 \otimes #2 \, (#3)}

\newcommand{\posPart}[1]{\left[#1\right]^+}

\newcommand{\verDev}[2]{v \! \left(#1, #2\right)}
\newcommand{\horDev}[2]{h \! \left(#1, #2\right)}

\newcommand{\fn}{\mathcal{F}^\uparrow_0}
\newcommand{\f}{\mathcal{F}}
\newcommand{\fa}{\mathcal{F}^\uparrow}
\newcommand{\ma}{\beta_{\mathrm{res}}^{\mathrm{mac}}}
\newcommand{\ca}{\beta_{\mathrm{res}}^{\mathrm{ca}}}
\newcommand{\cai}{\beta_{\mathrm{res}}^{\mathrm{ca},i}}
\newcommand{\madelay}{d_{\mathrm{e2e}}^{\mathrm{mac}}}
\newcommand{\cadelay}{d_{\mathrm{e2e}}^{\mathrm{ca}}}
\newcommand{\cat}{T_{\mathrm{res}}^{\mathrm{ca},i}}
\newcommand{\rat}{R_{\mathrm{res}}^{\mathrm{ca},i}}
\newcommand{\awc}{A^{\mathrm{WC}}}
\newcommand{\dwc}{D^{\mathrm{WC}}}
\newcommand{\tvd}{t_{\mathrm{vd}}}
\newcommand{\rlo}{R^{\mathrm{res}}}
\newcommand{\tlo}{T^{\mathrm{res}}}
\newcommand{\blo}{B^{\mathrm{res}}}
\newcommand{\rlac}{\underline{\beta}_{\underline{r}_1, T_{\underline{\alpha}_1}}}

\newcommand{\rl}{\beta_{R,T}}

\newcommand\numberthis{\stepcounter{equation}\tag{\theequation}}

\providecommand{\lemmaname}{Lemma}

\allowdisplaybreaks

\begin{document}

\title{
Extending Network Calculus To Deal With Partially Negative And Decreasing Service Curves
}

\author{\IEEEauthorblockN{
Anja Hamscher, Vlad-Cristian Constantin, Jens B. Schmitt}
\IEEEauthorblockA{
\textit{Distributed Computer Systems (DISCO) Lab} \\ 
\textit{RPTU Kaiserslautern-Landau, Germany}\\
\{hamscher,constantin,jschmitt\}@cs.uni-kl.de}
}

\maketitle

\begin{abstract}
Network Calculus (NC) is a versatile analytical methodology to efficiently compute performance bounds in networked systems.
The arrival and service curve abstractions allow to model diverse and heterogeneous distributed systems. The operations to compute residual service curves and to concatenate sequences of systems enable an efficient and accurate calculation of per-flow timing guarantees.
Yet, in some scenarios involving multiple concurrent flows at a system, the central notion of so-called \emph{min-plus} service curves is too weak to still be able to compute a meaningful residual service curve. 
In these cases, one usually resorts to so-called \emph{strict} service curves that enable the computation of per-flow bounds. However, strict service curves are restrictive: (1) there are service elements for which only min-plus service curves can be provided but not strict ones and (2) strict service curves generally have no concatenation property, i.e., a sequence of two strict systems does not yield a strict service curve. 
In this report, we extend NC to deal with systems only offering aggregate min-plus service curves to multiple flows. The key to this extension is the exploitation of \emph{minimal arrival curves}, i.e., lower bounds on the arrival process. 
Technically speaking, we provide basic performance bounds (backlog and delay) for the case of \emph{negative} service curves. We also discuss their accuracy and show them to be tight.
In order to illustrate their usefulness we also present patterns of application of these new results for: (1) heterogeneous systems involving computation and communication resources and (2) finite buffers that are shared between multiple flows.
\end{abstract}

\section{Introduction}

Network Calculus (NC) has proven to be a useful analytical methodology in the worst-case performance analysis of networked systems. As a stateless method it is computationally efficient and allows for extensive design space explorations. As such, it has seen numerous usage in real-world systems (e.g., TSN \cite{maile2020network, de2014complete, zhao2018worst}, AFDX \cite{charara2006methods, frances2006using, boyer2008tightening}, Network-on-Chip \cite{bakhouya2009analytical, boyer2020bounding}).

NC provides a rich set of results: it can deal with all kinds of arrival processes and service elements. Its strength lies in providing a (min, plus) system theory that enables a tight or at  least  accurate end-to-end delay analysis. It was pioneered by Cruz \cite{cruz1991calc1, cruz1991calc2} and Chang \cite{chang2000performance}, a comprehensive and up-to-date account of NC results is given in \cite{bouillard2018deterministic}. 
A central notion in NC is the service curve, abstracting scheduling disciplines at communication and computational resources. Several definitions exist, the two main ones being min-plus and strict service curves.
Strict service captures the system behavior in a relatively tight manner, whereas the min-plus service curve is a weaker approximation, but comes with nice mathematical properties.
Many different NC analysis methods, from Total Flow Analysis \cite{cruz1991calc2} over PMOO \cite{schmitt2008improving, schmitt2008delay} to Deep Tandem Matching Analysis \cite{geyer2019deeptma, bondorf2017quality}, have been developed over the years to accommodate for different system topologies and provide different trade-offs between accuracy of the bounds and computational cost.

\begin{figure}
    \centering
    \includegraphics[width=\columnwidth]{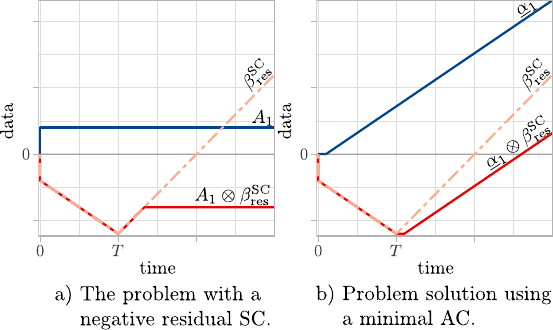}
    \caption{Problem statement and solution.}
    \label{fig:problem}
\end{figure}

Yet, there is a \emph{blind spot} of NC: when a residual per-flow service curve needs to be calculated from an aggregate \emph{min-plus} service curve (instead of a strict one). In this case, \cite[p.~161]{bouillard2018deterministic} states that it is not feasible to compute performance bounds because residual service curves become partially negative and decreasing. These service curve properties cause an issue with certain arrival patterns, when there is not enough input traffic to "drive the system forward". In other words, the server is not forced to serve any arrivals as its departure guarantee never becomes positive. The problem is illustrated in Fig.~\ref{fig:problem}a (from \cite{bouillard2018deterministic}). To avoid this problem, we propose the use of a \emph{minimal arrival curve}, i.e., a lower bound on the input to a system. Then, the departure guarantee at a server is determined by the minimal arrival curve, ensuring that the guarantee eventually becomes positive, as illustrated in Fig.~\ref{fig:problem}b. 

In conclusion, the key idea of this report is to use minimal arrival curves to enable a performance analysis using NC in multiple flow scenarios when strict service curves cannot be assumed, or more generally, when per-flow service curves can become partially negative and decreasing. Overall, we make the following contributions in this report:
\begin{itemize}
    \item We extend NC such that the calculation of performance bounds is also possible for partially negative min-plus service curves in Sect.~\ref{sec:bounds}. While this is completely novel for the delay bound, the conventional backlog bound remains largely the same with a slight adaptation.
    \item We discuss the accuracy of both delay and backlog bounds. For the delay bound, we show in Sect.~\ref{sec:delay} that it is tight. For the backlog bound, we show that it is also tight by providing a non-trivial sample path argument in Sect.~\ref{sec:backlog}. 
    \item We present patterns of application demonstrating the practical usefulness of the new results in Sect.~\ref{sec:apps}. In fact, in several cases the novel bounds outperform state-of-the-art techniques, or even enable an analysis at all.
\end{itemize}

\section{Background}\label{sec:background}
In this section, we introduce the necessary background on network calculus and recapitulate existing results regarding multiple flow scenarios. 

\subsection{Some Mathematical Background}

Let $a,b\in\mathbb{R}$. We call $\wedge$ the \emph{minimum operator} with $a\wedge b:=\min\{a,b\}$, and $\vee$ the \emph{maximum operator} with $a\vee b:=\max\{a,b\}$. The function $\posPart{a} \coloneqq \max\{0, a\}$ yields the \emph{positive part} of the argument $a$. 

We make use of certain properties of sets. Let $P,Q \subseteq \mathbb{R}$ be two non-empty sets of real numbers. It holds that
\begin{align}
\label{eq:minsupinf}
\begin{split}
 -\sup P=\inf P^- ,
 -\inf P = \sup P^- ,
\end{split}
\end{align}
where $P^-\coloneqq \left\{-x\mid x\in P \right\}.$
Infimum and supremum exhibit the following properties:
\begin{align}
\label{eq:split}
\begin{split}
 \sup (P\cup Q)=(\sup P) \vee (\sup Q) ,
\\
 \inf (P\cup Q)=(\inf P) \wedge (\inf Q) .
\end{split}
\end{align}
Moreover, if $P\subseteq Q$, it holds that 
\begin{equation}\label{eq:subset}
    \sup P \leq \sup Q,\  \inf P \geq \inf Q.
\end{equation}

\subsection{Network Calculus Background}

We begin by defining several function classes \cite[p.~22]{bouillard2018deterministic} that are used throughout the report. Let $\mathbb{R}^+$ be the set of non-negative real numbers. $\f:=\{f: \mathbb{R}^+ \rightarrow \mathbb{R}\cup\{+\infty\}\}$ is the set of (min, plus) functions. Based on $\f$, we let $\fa$ be the set of non-decreasing functions $f\in\mathcal{F}$, and $\fn$ be the set of functions in $\fa$ with $f(0)=0$. Similarly, we introduce the following sets: $\mathcal{F}^+_{<0}$ is the set of functions in $\fa$ with $f(0)<0$ and $\mathcal{F}^{\uparrow}_{\leq0}$ is the set of functions in $\fa$ with $f(0)\leq0$. 

\begin{defn}
A function $f\in\f$ is right-continuous if $\forall t \in \mathbb{R}$,
\[
f(t^+) \coloneqq \lim_{s \searrow t} f(s) \coloneqq \lim_{s \to t, s > t} f(s)
\]
always exists and is equal to $f(t)$.
\end{defn} 

\begin{defn}[Pseudo-inverse]\label{def:pseudo}
Let $f\in\mathcal{F}^{\uparrow}$ be a non-negative and non-decreasing function. Then, the pseudo-inverse $f^{-1}$ is defined $\forall x\geq0$ as
\begin{equation}\label{eq:pseudo}
    f^{-1}(x)=\inf\left\{t\mid f(t)\geq x\right\}.
\end{equation}
\end{defn}

\begin{defn}[Shift Function]\label{eq:delta-function}
	The \emph{shift function} is defined by
	\begin{equation}
	\delta_T(t) \coloneqq \begin{cases*}
	+\infty, & if $ t>T ,$\\
	0, & otherwise.
	\end{cases*}
	\label{eq:delta-function}
	\end{equation}
\end{defn}

\begin{defn}[Operators \cite{bouillard2018deterministic}]\label{def:operators}
Let $f, g\in\f$ be two functions. The \emph{(min, plus) convolution} of $f$ and $g$ is defined as $f\otimes g(t)\coloneqq \inf_{0\leq s\leq t}\{f(t-s) + g(s)\}$, the \emph{(min, plus) deconvolution} is defined as $f\oslash g(t)\coloneqq \sup_{s\geq 0} \{f(t+s) - g(s)\}$. The \emph{(max, plus) convolution} is defined as $f\up{\otimes}g(t)\coloneqq \sup_{0\leq s\leq t}\{f(t-s)+g(s)\}$, and the \emph{(max, plus) deconvolution} is defined as $f\up{\oslash}g(t)\coloneqq \inf_{s\geq 0}\{f(t+s)-g(s)\}$.

\end{defn}
We introduce several properties of these operators.
\begin{rem}[Isotonicity of $\otimes$\cite{le2001network}]\label{rem:conv}
Let $f,g,f',g'\in\f$. If $f\leq g$ and $f'\leq g'$, then $f\otimes f'\leq g\otimes g'$.
\end{rem}

\begin{prop}[Composition of $\oslash$ and $\otimes$]
	\label{prop:composition-deconv-conv}
	Let $ f,g,h \in \mathcal{F}:$
\begin{equation}\label{eq:composition_prop}
    (f \otimes g) \oslash h \leq f \otimes (g \oslash h).
\end{equation}
\end{prop}

Next, we define various notions that are used to model a network and derive its performance bounds. Let $A, D\in\fn$ be the \emph{cumulative arrival} and \emph{departure process} of a flow in the network, assuming causality $A\geq D$. Furthermore, we assume all systems to be lossless. We define the most important performance measures for such a system:
\begin{defn}[Backlog at Time $t$]\label{def:backlog}
	The \emph{backlog} of system $\mathcal{S}$ at time $t$ is the vertical
	distance between arrival process $A$ and departure process $D$
	at time $t$,
	\begin{equation}
	q(t) \coloneqq A(t)-D(t).
	\label{eq:backlog}
	\end{equation}
\end{defn}

\begin{defn}[Virtual Delay at Time $t$]
	\label{def:delay}
	The \emph{virtual delay} of data arriving at system $\mathcal{S}$ at time $t$ is the time until this data would be served, assuming FIFO order of service,
	\begin{equation}
	d(t) \coloneqq \inf\left\{ \tau\geq 0: A(t)\leq D(t+\tau)\right\}.
	\label{eq:virtual-delay}
	\end{equation}
\end{defn}

Arrival and service curves are essential elements of the performance analysis using NC. We define arrival curves first.
\begin{defn}[Maximal and Minimal Arrival Curve]
	\label{def:arrival-curve}
	Let $\up{\alpha}, \mac{\alpha}\in\fn$. We say that $\up{\alpha}$ is a \emph{maximal arrival curve} for arrival process $A$, and $\mac{\alpha}$ is a \emph{minimal arrival curve} for $A$, if it holds for all $0\leq s\leq t$ that
	\begin{equation}
		\mac{\alpha}(t-s) \leq A(t) - A(s) \leq \up{\alpha}(t-s). 
	\end{equation}\label{ineq:ac}

\end{defn}
\vspace*{-6mm}
A frequent example is the \emph{token-bucket} arrival curve $ \gamma_{r,b}(t) = b+rt $ for $t > 0$, $\gamma_{r,b}(0)=0$. Note that $\gamma_{r_1,b_1}+\gamma_{r_2,b_2}=\gamma_{r_1+r_2, b_1+b_2}$. Next, we define service curves.

\begin{defn}[Service Curve (SC)]\label{def:service-curve}
    Let a flow with arrival process $A$ and departure process $D$ traverse a system $\mathcal{S}$. The system offers a \emph{min-plus service curve} $\beta$ to the flow if $\beta \in \f$ and it holds for all $t \geq 0$ that \vspace*{-1mm}
	\begin{equation}\label{ineq:servicecurve}
		D(t) \geq \conv{A}{\beta}{t} = \inf_{0\leq s\leq t} \left\{A(t-s) + \beta(s)\right\}.
	\end{equation}
\end{defn}\vspace*{-1mm}

Often, $\beta\in\fn$ is assumed, yet we let $\beta\in\f$ as in \cite{bouillard2018deterministic}.

\begin{defn}[Strict Service Curve (SSC)]
	\label{def:strict-service-curve}
	A system offers a \emph{strict service curve} $\beta \in \mathcal{F}$ to a flow if, during any backlogged period $(s,t]$ (i.e. $\forall t'\in (s,t], q(t')>0$), it holds that
	\begin{equation}\label{ineq:strict-service-curve}
		D(t) - D(s) \geq \beta(t-s).
	\end{equation}
\end{defn}

A frequently employed function for minimal arrival and service curves is the \emph{rate-latency} curve $ \beta_{R,T}(t) \coloneqq R\cdot \posPart{t-T}$.

We define two characteristic distances between functions.
\begin{defn}
	Let $f,g\in\f$. The \emph{vertical deviation} between $f$ and
	$g$ is defined as
	\begin{equation}\label{eq:vertical-deviation}
		\verDev{f}{g} \coloneqq \sup_{t\geq0}\left\{ f(t)-g(t)\right\},
	\end{equation}
	and the \emph{horizontal deviation} between $f$ and $g$ is defined as
	\begin{align*}
	\horDev{f}{g} &\coloneqq \sup_{t\geq0}\left\{ \inf\left\{ \tau \geq 0 \mid f(t)\leq g(t+\tau)\right\}\right\}\numberthis \label{eq:horizontal-deviation}\\
	&=\inf\left\{ \tau \geq 0 \mid \sup_{t \geq 0} \left\{f(t) - g(t+\tau)\right\} \leq 0\right\}\numberthis \label{eq:alt-horizontal-deviation}.
	\end{align*}
\end{defn}

There is a useful property of deviations \cite[p.~115]{bouillard2018deterministic}:
\begin{lem}[Monotony of Deviations]\label{lem:mon_dev}
For all $f,f',g,g'\in\fa$, if $f\geq f'$ and $g\leq g'$, then
\begin{equation}\label{ineq:mon_dev}
    \verDev{f}{g}\geq\verDev{f'}{g'} 
    \quad\text{and}\quad
    \horDev{f}{g}\geq\horDev{f'}{g'}. 
\end{equation}
\end{lem}

Using these concepts, one can derive performance bounds for the measures defined previously\cite[p.~115]{bouillard2018deterministic},\cite[p.~118]{le2001network}. 

\begin{thm}[Performance Bounds]\label{thm:performancebounds}
    Assume an arrival process $A$, constrained by maximal arrival curve $\up{\alpha}\in\fn$, traverses a system $\mathcal{S}$. Let the system $\mathcal{S}$ offer a service curve $\beta\in\fn$. The virtual delay $d(t)$ satisfies for all t \vspace*{-1mm}
    \begin{equation}\label{eq:delaybound}
        d(t)\leq \horDev{\up{\alpha}}{\beta}.
    \end{equation}
    The  backlog $q(t)$ satisfies for all t  
    \begin{equation}\label{eq:backlogbound}
        q(t)\leq \verDev{\up{\alpha}}{\beta}. 
    \end{equation}
\end{thm}
Note that Thm.~\ref{thm:performancebounds} requires $\beta\in\fn$. 

We can also calculate a bound on the departure process $D$ of a system offering a min-plus service curve $\beta \in \f$:
\begin{equation}\label{eq:outputbound}
    D\leq \up{\alpha}\oslash\beta .
\end{equation} 

A central result of NC is the concatenation theorem.
\begin{thm}[Concatenation Theorem]\label{def:concat}
Let a flow with arrival process $A$ traverse systems $\mathcal{S}_1$ and $\mathcal{S}_2$, offering service curves $\beta_1, \beta_2\in\f$, in sequence. Then, the concatenation of the two systems, $S_{1,2}=\langle S_1, S_2 \rangle,$ offers an end-to-end service curve $\beta_{1,2}=\beta_1\otimes\beta_2$ to the arrival process. 
\end{thm}

\begin{defn}[Sub-additive and Super-additive Functions]
    \label{def:subadditive}
    Let $f \in \mathcal{F}$.
    Then $f$ is \emph{sub-additive} if for all $ s,t \geq 0 $
    \begin{equation}
    f(t+s)\leq f(t)+f(s).
    \label{ineq:subadditive}
    \end{equation}
    On the other hand, $f$ is \emph{super-additive} if for all $ s,t \geq 0 $
    \begin{equation}
		f(t+s)\geq f(t)+f(s).
    \label{ineq:superadditive}
    \end{equation}
\end{defn}

\begin{prop}\label{rem:subadd}
Let $f \in \fn$ and be sub-additive. If $f\neq0$ then $\forall t > 0: f(t)>0$. 
\end{prop}
\begin{proof}
We prove this by contradiction:
assume $\exists t_0>0$ with $f(t_0)=0$. Since $f$ is non-decreasing, then $f(t)=0 \; \forall t\in[0,t_0]$. For any $t_1>t_0$, due to the sub-additivity of $f$ (and being non-decreasing), it holds that 
\begin{equation*}
    f(t_1)\leq \left\lceil \frac{t_1}{t_0} \right\rceil f(t_0) =0
\end{equation*}
This means in turn that $f=0$, which contradicts the assumption that $f\neq0$ and thus $\forall t_0 > 0: f(t_0)>0$.
\end{proof}

\begin{defn}[Sub-additive Closure \cite{le2001network}]
\label{def:subadd-closure}
Let $f\in\f$.
The sub-additive closure of $f$ is defined by
\begin{equation}
    f^* \coloneqq \inf_{n\geq 0}\left\{f^{(n)}\right\},
\end{equation}
where $f^{(n)}$ is the $n$-fold self-convolution of $f$, i.e., $ f^{(0)} = \delta_0 $, $ f^{(1)} = f $ and $f^{(n)}=\bigotimes_{i=0}^{n}f^{(i)}$ for $n\geq 2.$
\end{defn}

With respect to tightness, we remark that maximal arrival curves that are not sub-additive and minimal arrival curves that are not super-additive can be improved by replacing them by their sub-additive and super-additive closures, respectively (see \cite{bouillard2018deterministic}, Propositions 5.2 and 5.3). 

Moreover. we also note that both arrival curves may be further improved by combining their respective information \cite{moy2010arrival} (see also 
\cite[Theorem~5.1]{bouillard2018deterministic}).

\section{Extension of NC Performance Bounds for Negative Service Curves}\label{sec:bounds}

In this section, we extend the performance bounds presented in Thm.~\ref{thm:performancebounds} to the case where the service curve $\beta \notin \fn$. 

For the service curve under consideration we make the very general assumption that $\beta\in\f_{\leq 0}$. Note that $\beta(0)\leq 0$ means no loss of generality \cite[p.~107]{bouillard2018deterministic}. 

Next, in order to be able to focus on the negativity of service curves, we "safely" replace the original service curve $\beta \in\mathcal{F}$ by $\xi= \beta_\downarrow:=\beta \up{\oslash}0$ \cite[p.~107]{bouillard2018deterministic}.
$\beta_\downarrow$ is the largest non-decreasing function with $\beta_\downarrow\leq \beta$, which is why we call it the \emph{lower} non-decreasing closure. Then, by isotonicity of the (min, plus) convolution (see Remark \ref{rem:conv}), $\xi$ is also a service curve. Note that this is different from the (upper) non-decreasing closure as defined in \cite[p.~45]{bouillard2018deterministic}. 
While the lower non-decreasing closure is safe to use as $\xi\leq\beta$, it might be conservative. 

It is clear that
$\text{\ensuremath{\xi\in\mathcal{\mathcal{F_{\leq\text{0}}^{\uparrow}}}}}$.
In particular, $\xi\in\mathcal{F}_{0}^{\uparrow}$ if and only if $\beta\geq0$,
and $\xi\in\mathcal{F}_{<0}^{\uparrow}$ if and only if $\exists s\geq0$ with
$\beta(s)<0$.

\subsection{Generalizing the Delay Bound} \label{sec:delay}

We start with generalizing the delay bound and discuss its tightness thereafter. Let us first state a useful technical lemma.

\begin{lem}\label{lem:delay-bound}
Let $f,g \in\f$ be non-increasing. Then,
\begin{align*}
    &\inf\left\{ \tau\geq0\mid f(\tau)\leq0\right\}\vee \inf\left\{ \tau\geq0\mid g(\tau)\leq0\right\}\\
    &=\inf\left\{ \tau\geq0\mid (f\vee g)(\tau)\leq0\right\}.\numberthis \label{eq:lem_dbound}
\end{align*}
\end{lem}
\begin{proof}
Let $\tau_{f}\coloneqq\arg\inf\left\{ \tau\geq0\mid f(\tau)\leq0\right\}$. We define $\tau_{g}$ 
similarly. 
Note that for $f>0$, $\tau_{f}=\infty$. For the moment, let $f$ be right-continuous at $\tau_f$, then we have $f(\tau)\leq0,\forall\tau\geq\tau_{f}$, since $f$ is non-increasing. This clearly also holds for $g$. Then,
\begin{align*}
    &\inf\left\{ \tau\geq0\mid f(\tau)\leq0\right\}\vee \inf\left\{ \tau\geq0\mid g(\tau)\leq0\right\}\\
    &=\inf[\tau_{f},+\infty)\vee\inf[\tau_{g},+\infty)\\
    &=\inf[\tau_{f}\vee\tau_{g},+\infty)\\
    &=\inf[\tau_{f},+\infty)\cap[\tau_{g},+\infty)\\
    &= \inf\left\{ \tau\geq0\mid f(\tau)\leq0\right\}\cap\left\{ \tau\geq0\mid g(\tau)\leq0\right\}\\
    &=\inf\left\{ \tau\geq0\mid f(\tau)\leq0\textit{ AND } g(\tau)\leq0\right\}\\
    &=\inf\left\{ \tau\geq0\mid (f\vee g)(\tau)\leq0\right\}.
\end{align*}

Now, in case $f$ is not right-continuous at $\tau_f$, we have that $\left\{ \tau\geq0\mid f(\tau)\leq0\right\}=(\tau_{f},+\infty)$, and the proof proceeds along the same lines.
\end{proof}

\begin{thm}[Generalized Delay Bound]
\label{thm:delay-bound-1} Let an arrival process $A$ traverse
a system $\mathcal{S}$. Further, let the arrivals be constrained
by maximal arrival curve $\up{\alpha}\in\fn$ and minimal arrival curve $\mac{\alpha}\in\fn$,
and let the system offer a service curve $\xi\in\f_{\leq0}^{\uparrow}$
.The virtual delay $d(t)$ satisfies for all $t\geq0$
\begin{equation}
d(t)\leq z(\mac{\alpha},\xi)\vee h(\up{\alpha},\xi),\label{ineq:delay-bound-1}
\end{equation}
with $z(\mac{\alpha}, \xi)\coloneqq\inf\left\{ \tau\geq0\mid\mac{\alpha}\otimes\xi(\tau)\geq0\right\}$.
\end{thm}

\begin{proof} 
First, consider the case when $\xi\in\mathcal{F_{\text{0}}^{\uparrow}}$.
This is the classical case from Thm.~\ref{thm:performancebounds}, for which
we know that $d(t)\leq h(\up{\alpha},\xi)$. It suffices to show that
$d(t)\leq z(\mac{\alpha},\xi)\vee h(\up{\alpha},\xi)$.
We have that
\begin{equation}
    z(\mac{\alpha},\xi)=\inf\left\{ \tau\geq0\mid\mac{\alpha}\otimes\xi(\tau)\geq0\right\} =0,
\end{equation}
since $\mac{\alpha}(0)=\xi(0)=0$. Therefore,
\[
d(t)\leq z(\mac{\alpha},\xi)\vee h(\up{\alpha},\xi)=h(\up{\alpha},\xi).
\]

Next, consider the case when $\xi\in\mathcal{F_{<\text{0}}^{\uparrow}}$. We derive 
\begin{align*}
        d(t)\overset{\eqref{eq:virtual-delay}}{=} & \inf\left\{ \tau\geq0\mid D(t+\tau)\geq A(t)\right\} \\
        \overset{\eqref{ineq:servicecurve},\eqref{eq:subset}}{\leq} & \inf\left\{ \tau\geq0\mid A\otimes\xi(t+\tau)\geq A(t)\right\} \\
        = &
        \inf\left\{ \tau\geq0\mid\inf_{0\leq s\leq t+\tau}\left\{ A(t+\tau-s)+\xi(s)\right\} \geq A(t)\right\} \\
        = & \inf\left\{ \tau\geq0\mid A(t)-\inf_{0\leq s\leq t+\tau}\left\{ A(t-(s-\tau))+\xi(s)\right\}\right. \\ &\left.\hphantom{\tau\geq0\mid A(t)-\inf_{0\leq s\leq t+\tau}\left\{ A(t-(s-\tau))+\xi ss \right\}}\leq0\vphantom{\inf_{0\leq s\leq t+\tau}\left\{ A(t-(s-\tau))+\xi(s)\right\}}\right\} \\ 
        \overset{\eqref{eq:minsupinf}}{=} & \inf\left\{ \tau\geq0\mid\sup_{0\leq s\leq t+\tau}\left\{ A(t)-A(t-(s-\tau))-\xi(s)\right\}\right. \\ &\left. \hphantom{\tau\geq0\mid A(t)-\inf_{0\leq s\leq t+\tau}\left\{A(t-(s-\tau))+\xi ss\right\}}\leq0\vphantom{\sup_{0\leq s\leq t+\tau}\left\{ A(t)-A(t-(s-\tau))-\xi(s)\right\}}\right\} \\
        \overset{\eqref{eq:split}}{=} & \inf\left\{ \tau\geq0\mid\sup_{0\leq s\leq\tau}\left\{ A(t)-A(t-(s-\tau))-\xi(s)\right\}\right. \\ &\left. \vee  
        \sup_{\tau<s\leq t+\tau}
        \left\{ A(t)-A(t-(s-\tau))-\xi(s)\right\} \leq0\right\}
        \\
        \overset{\eqref{ineq:ac},\eqref{eq:subset}}{\leq} & \inf\left\{ \tau\geq0\mid\sup_{0\leq s\leq\tau}\left\{ -\mac{\alpha}(\tau-s)-\xi(s)\right\}\right. \\ &\left.  \vee\sup_{\tau<s\leq t+\tau}\left\{ \up{\alpha}(s-\tau)-\xi(s)\right\} \leq0\right\} 
        \\
        \overset{\eqref{eq:minsupinf}}{=}& \inf\left\{ \tau\geq0\mid-\mac{\alpha}\otimes\xi(\tau)  \vee\sup_{0<s'\leq t}\left\{ \up{\alpha}(s')-\xi(s'+\tau)\right\} \right. \\ &\left. \hphantom{\tau\geq0\mid A(t)-\inf_{0\leq s\leq t+\tau}\left\{A(t-(s-\tau))\right\}}\leq0\vphantom{\sup_{0\leq s\leq t+\tau}}\right\} \hspace*{-4mm}\numberthis \label{dbound:new_step}\\
        \overset{\eqref{eq:lem_dbound}}{=} & \inf\left\{ \tau\geq0\mid-\mac{\alpha}\otimes\xi(\tau) \leq0\right\} \\ & \vee\inf\left\{ \tau\geq0\mid\sup_{0<s'\leq t}\left\{ \up{\alpha}(s')-\xi(s'+\tau)\right\} \leq0\right\} 
        \\
        \overset{\eqref{eq:subset}}{\leq} & \inf\left\{ \tau\geq0\mid\mac{\alpha}\otimes\xi(\tau)\geq0\right\}\\ & \vee\sup_{s'\geq0}\left\{\inf\left\{ \tau\geq0\mid \up{\alpha}(s')-\xi(s'+\tau) \leq0\right\}\right\} 
        \numberthis \label{dbound:step9}
        \\
        = & z(\mac{\alpha},\xi)\vee h(\up{\alpha},\xi).
\end{align*}

In line 8 (Eq.~\eqref{dbound:new_step}) we make the substitution $s'\coloneqq s-\tau $.  In line 10 (Eq.~\eqref{dbound:step9}) we rewrite the supremum as in Eq.~\eqref{eq:alt-horizontal-deviation} and take the supremum over a larger set.
It is left to check that the conditions of Lem.~\ref{lem:delay-bound} in line 8 (Eq.~\eqref{dbound:new_step}) apply:
\begin{itemize}
    \item due to the closedness 
 of the (min, plus) convolution for the set of non-decreasing functions \cite[p.~22]{bouillard2018deterministic}, and both $\mac{\alpha}$
and $\xi$ being non-decreasing, we see that $-\mac{\alpha}\otimes\xi(\tau)$ is non-increasing in $\tau$;
    \item since $\xi$ is non-decreasing, $\sup_{0<s\leq t}\left\{ \up{\alpha}(s)-\xi(s+\tau)\right\}$ is clearly non-increasing in $\tau$.
\end{itemize}
\end{proof}

\begin{figure}
    \centering
    \includegraphics[width=0.6\columnwidth]{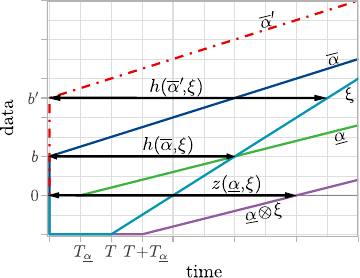}
    \caption{Different cases of the generalized delay bound theorem.}
    \label{fig:delaycases}
\end{figure}

For an illustrative example, showing the different cases governing the generalized delay bound, see Fig.~\ref{fig:delaycases}. Here, we 
 assume a maximal token-bucket arrival curve $\up{\alpha}=\gamma_{r,b}$, a minimal rate-latency arrival curve $\mac{\alpha}=\beta_{R,T_{\mac{\alpha}}}$ and a simple negative service $\xi$. We show two cases of maximal arrival curves with different burst sizes such that both cases ($h(\up{\alpha},\xi)$ and $z(\mac{\alpha},\xi)$) of the generalized delay bound are provoked. One can observe that in case of $\up{\alpha}$ having a smaller burst, the delay bound is given by $z(\mac{\alpha},\xi)$, whereas when we have a burstier maximal arrival curve $\up{\alpha}'$ then $h(\up{\alpha}',\xi)$ dominates. 

So, we have extended the delay bound analysis to functions which are not in $\fn$. We show below that, under mild assumptions, the new delay bound is actually tight. Before that we provide a helpful lemma. 

\begin{lem}[$z$-Monotony]\label{lem:z_monotony}
For all $f,f',g,g'\in\fa$, if $f\leq f'$ and $g\leq g'$, then
\begin{equation}\label{ineq:z_monotony}
    z(f,g)\geq z(f',g'). 
\end{equation}
\end{lem}
\begin{proof}
Set $Z(f,g)\coloneqq\left\{ \tau\geq0\mid f\otimes g(\tau)\geq0\right\}$. We define $Z(f,g')$ and $Z(f',g')$ similarly. We then have $\forall \tau\geq0$
\[f\otimes g(\tau)\geq0\implies f\otimes g'(\tau)\geq0\implies f'\otimes g'(\tau)\geq0,\]
since $f\otimes g\leq f\otimes g'\leq f'\otimes g'$. This implies that
\begin{equation*}
    Z(f,g)\subseteq Z(f,g')\subseteq Z(f',g'),
\end{equation*}
which in turn means 
\begin{equation*}
    z(f,g)=\inf Z(f,g)\geq\inf Z(f',g')=z(f',g'),
\end{equation*}
by Property~\eqref{eq:subset}. 
\end{proof}

\begin{thm}[Tightness of the Generalized Delay Bound]
\label{thm:delay-bound-tightness-alternative} Let an arrival
process $A$ traverse a system $\mathcal{S}$. Further, let the
arrivals be constrained by a sub-additive maximal arrival curve $\up{\alpha}\in\fn$ and  a super-additive minimal arrival
curve\textup{ $\mac{\alpha}\in\fn$}, and assume these cannot be further improved by combining their respective information (see \cite[Theorem~5.1]{bouillard2018deterministic}).
Let the system offer a
service curve $\xi\in\f_{\leq0}^{\uparrow}$. We also assume that $\mac{\alpha}$
and $\xi$ are right-continuous and that $\up{\alpha}(0_+)>0$, i.e., the arrivals exhibit a non-zero burstiness.
\\
If $h(\up{\alpha},\xi)> z(\mac{\alpha},\xi)$, we set  \textup{$\awc \coloneqq\up{\alpha}$} and 
$\dwc\coloneqq\left[\up{\alpha}\otimes\xi\right]^{+}$, then the worst-case delay ($\mathrm{WCD}$) is
\begin{equation}
\mathrm{WCD}=h(\awc,\dwc)=h(\up{\alpha},\xi).
\end{equation}

If $h(\up{\alpha},\xi) \leq z(\mac{\alpha},\xi)$, we set  \textup{$\awc(t) \coloneqq\mac{\alpha}(t)+\up{\alpha}(0_+)\cdot\mathbbm{1}_{\{t>0\}}$} and 
$\dwc(t)\coloneqq\left[\mac{\alpha}\otimes\xi(t)+\up{\alpha}(0_+)\cdot\mathbbm{1}_{\{t>0\}}\right]^{+}$ for all $t\geq0$, then the worst-case delay ($\mathrm{WCD}$) is
\begin{equation}
WCD=d(0_+)=z(\mac{\alpha},\xi),
\end{equation}
for $d(0_+)\coloneqq\displaystyle\lim_{\substack{t\rightarrow0 \\ t>0}}d(t)$ and $\up{\alpha}(0_+)\coloneqq\displaystyle\lim_{\substack{t\rightarrow0 \\ t>0}}\up{\alpha}(t)$.
\end{thm}

\begin{proof}
We start with the first case, where $h(\up{\alpha},\xi)> z(\mac{\alpha},\xi)$, and \textup{$\awc\coloneqq\up{\alpha}$ as well as} $\dwc\coloneqq\left[\up{\alpha}\otimes\xi\right]^{+}$. Then
\begin{align*}
h(\awc,\dwc)=&h(\up{\alpha},\left[\up{\alpha}\otimes\xi\right]^{+}) \\
\overset{\eqref{eq:horizontal-deviation}}{=}& \sup_{t\geq0}\left\{ \inf\left\{ \tau\geq0\mid \up{\alpha}(t)\leq\left[\up{\alpha}\otimes\xi(t+\tau)\right]^{+}
\right.\right. \\ &\left.\left.\hphantom{\sup_{t\geq0}\{ \inf\{ \tau\geq0\mid \up{\alpha}(t)\leq\left[\up{\alpha}\otimes\xi(t+\tau)\right]}
\vphantom{\inf\left\{ \tau\geq0\mid \up{\alpha}(t)\leq\left[\up{\alpha}\otimes\xi(t+\tau)\right]^{+}\right\}}\right\} \right\} \\
\overset{\eqref{eq:split}}{=} & \inf\{\tau\geq0\mid \up{\alpha}(0)\leq[\up{\alpha}\otimes\xi(\tau)]^+\}\\ 
&\vee\sup_{t>0}\left\{ \inf\left\{ \tau\geq0\mid \up{\alpha}(t)\leq[\up{\alpha}\otimes\xi(t+\tau)]^{+}
\right.\right. \\ &\left.\left.\hphantom{\sup_{t\geq0}\{ \inf\{ \tau\geq0\mid \up{\alpha}(t)\leq\left[\up{\alpha}\otimes\xi(t+\tau)\right]}
\right\} \right\}
\\
=&0\vee  \sup_{t>0}\left\{ \inf\left\{ \tau\geq0\mid \up{\alpha}(t)\leq
\right.\right. \\ &\left.\left.\hphantom{\sup_{t\geq0}\{ \inf\{ \tau\geq0\mid (t)\leq ..}
[\up{\alpha}\otimes\xi(t+\tau)]^{+}\right\} \right\}\\
=& \sup_{t>0}\left\{ \inf\left\{ \tau\geq0\mid \up{\alpha}(t)\leq\up{\alpha}\otimes\xi(t+\tau)\right\} \right\} \numberthis \label{tight_hdev:step5}\\
\overset{\eqref{eq:subset}}{\geq}&\sup_{t>0}\left\{ \inf\left\{ \tau\geq0\mid \up{\alpha}(t)\leq\xi(t+\tau)\right\} \right\}\numberthis \label{tight_hdev:step6}\\
=&h(\up{\alpha},\xi).
\end{align*}
In the fifth step (Eq.~\eqref{tight_hdev:step5}), due to the sub-additivity of $\up{\alpha}$, $\up{\alpha}(t)>0,\forall t>0$ (see Prop.~\ref{rem:subadd}), and thus the positive part becomes irrelevant. In the second to last line (Eq.~\eqref{tight_hdev:step6}) we use that $\up{\alpha}\otimes\xi\leq\delta_0\otimes\xi=\xi$, since $\up{\alpha}(0)=0$ and the convolution being isotonic (see Remark~\ref{rem:conv}). In the last line we take advantage of the observation that 
\begin{align*}
\inf\{\tau\geq0\mid \up{\alpha}(0)\leq\xi(\tau)\} =& \inf\{\tau\geq0\mid 0\leq\xi(\tau)\}\\
=& \inf\{\tau\geq0\mid 0\leq\delta_0\otimes\xi(\tau)\}\\
=& z(\delta_0,\xi) \\
\overset{\eqref{ineq:z_monotony}}{\leq}& z(\mac{\alpha},\xi)\\
< & h(\up{\alpha},\xi)
\end{align*}
This in turn means that the horizontal deviation is not taken at $t=0$, i.e.
\[h(\up{\alpha},\xi)=\sup_{t>0}\left\{ \inf\left\{ \tau\geq0\mid \up{\alpha}(t)\leq\xi(t+\tau)\right\} \right\}.\]

Let us consider now the second case, where  $h(\up{\alpha},\xi)\leq z(\mac{\alpha},\xi)$, and \textup{$\awc(t) \coloneqq\mac{\alpha}(t)+\up{\alpha}(0_+)\cdot\mathbbm{1}_{\{t>0\}}$} and 
$\dwc(t)\coloneqq\left[\mac{\alpha}\otimes\xi(t)+\up{\alpha}(0_+)\cdot\mathbbm{1}_{\{t>0\}}\right]^{+}$. Then, we have $\forall t>0$
\begin{align*}
d(t) = & \inf\left\{ \tau\geq0\mid \awc(t)\leq \dwc(t+\tau)\right\} \\
= &  \inf\left\{ \tau\geq0\mid \mac{\alpha}(t)+\up{\alpha}(0_+)\leq \left[\mac{\alpha}\otimes\xi(t+\tau)+\up{\alpha}(0_+)\right]^{+}\right\} \\
\geq &  \inf\left\{ \tau\geq0\mid \mac{\alpha}(t)+\up{\alpha}(0_+)\leq \left[\mac{\alpha}\otimes\xi(t+\tau)\right]^{+}+\up{\alpha}(0_+)\right\} \\
= &  \inf\left\{ \tau\geq0\mid \mac{\alpha}(t)+\up{\alpha}(0_+)\leq \mac{\alpha}\otimes\xi(t+\tau)+\up{\alpha}(0_+)\right\} \\
= &  \inf\left\{ \tau\geq0\mid \mac{\alpha}(t)\leq \mac{\alpha}\otimes\xi(t+\tau)\right\}. \numberthis \label{tight_hdev:step_before_lim}
\end{align*}
In the third line, we used the fact that $[f(t)+\up{\alpha}(0_+)]^+\leq[f(t)]^+ +\up{\alpha}(0_+)$. In the next line, due to $t>0$, we know that $\mac{\alpha}(t)+\up{\alpha}(0_+) >0$. 
Hence, for the condition within the infimum to hold, the positive part becomes irrelevant.\par
Now, since the lower bound from above (Eq.~\eqref{tight_hdev:step_before_lim}) holds $\forall t>0$, it also holds in the limit:
\begin{align*}
d(0_+)\geq& \inf\left\{ \tau\geq0\mid \displaystyle\lim_{\substack{t\rightarrow0 \\ t>0}}\mac{\alpha}(t)\leq \displaystyle\lim_{\substack{t\rightarrow0 \\ t>0}}\left(\mac{\alpha}\otimes\xi(t+\tau)\right)\right\} \\
= & \inf\left\{ \tau\geq0\mid 0 \leq \mac{\alpha}\otimes\xi(\tau)\right\} \\
= & z(\mac{\alpha},\xi).
\end{align*}
In the second line, we used the right-continuity of $\mac{\alpha}$ and of $\mac{\alpha}\otimes\xi$ \cite{guidolin2022looking}.

By Th. \ref{thm:delay-bound-1}, we have that $z(\mac{\alpha},\xi)$ is an upper-bound on the virtual delay and, thus, the generalized delay bound is tight.

Moreover, the created sample paths ($\awc$,$\dwc$) are conforming to their arrival and service curves. The system is also causal, i.e. $\awc\geq\dwc$, since $\xi(0)\leq0$ and, thus, for instance $\dwc = \posPart{\awc\otimes\xi} \leq \posPart{\awc} = \awc.$
\end{proof}
\begin{figure}
    \centering
    \includegraphics[width=0.6\columnwidth]{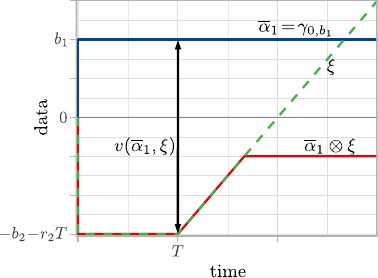}
    \caption{Vertical deviation in case of a negative service curve.}
    \label{fig:backlog}
\end{figure}

\begin{figure*}
    \centering
    \includegraphics[width=0.9\textwidth]{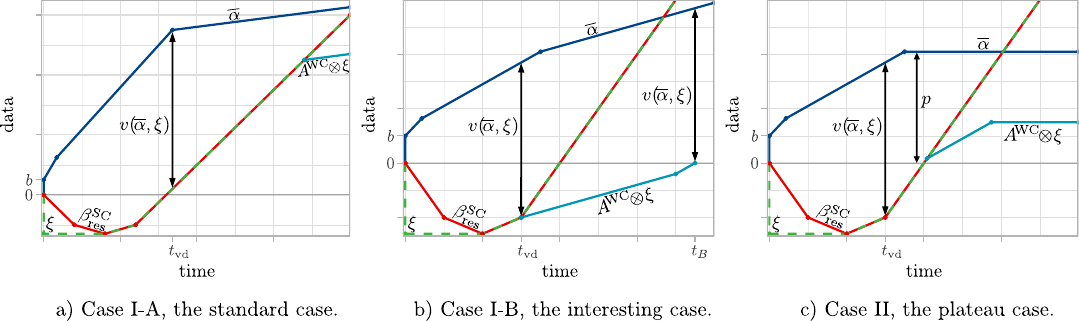}
    \caption{Graphical illustration of different cases in Thm.~\ref{thm:backlog-bound-tightness}.}
    \label{fig:th19_cases}
\end{figure*}

\subsection{Backlog Bound}\label{sec:backlog} 

While it is explicitly mentioned in \cite[p.~115]{bouillard2018deterministic} that service curves have to be an element of $\fa_0$ for finite delay bounds to exist, the same assumption is implicitly made for the backlog bound. However, as we show in the following, the backlog bound from Thm.~\ref{thm:performancebounds} can be applied to negative service curves with a slight technical adaptation and, more importantly, without the need to assume a minimal arrival curve. The latter becomes clear when looking at Fig.~\ref{fig:backlog}. We can see that the backlog remains finite even for this notorious example of a maximal arrival curve and thus without the departure guarantee ever becoming positive.

\begin{thm}[Backlog Bound]
\label{thm:backlog-bound} Let an arrival process $A$ traverse
a system $\mathcal{S}.$ Further, let the arrivals be constrained
by maximal arrival curve $\up{\alpha}\in\fn$, and let the system offer a service
curve $\xi\in\mathcal{F}_{\leq\text{0}}^{\uparrow}$. 
The backlog $q(t)$ satisfies for all $t$ 
\begin{equation}
q(t)\leq \verDev{\up{\alpha}}{\xi} \wedge\sup_{s\geq 0}\left\{ \up{\alpha}(s)\right\}.\label{ineq:backlog-bound}
\end{equation}
\end{thm}
\begin{proof}
	We have that
	\begin{align*}
		q(t) \overset{\eqref{eq:backlog}}{=}& A(t)-D(t) \\
		\overset{\eqref{ineq:servicecurve}}{\leq} &  A(t)-\conv{A}{\xi}{t} \\
		\overset{\eqref{eq:minsupinf}}{=}
		 & \sup_{0\leq s\leq t}\left\{ A(t)-A(t-s)-\xi(s)\right\} \\
		\overset{\eqref{ineq:ac}}{\leq} & \sup_{0\leq s\leq t}\left\{ \up{\alpha}(s)-\xi(s)\right\} \\
		\overset{\eqref{eq:subset}} 
		{\leq} & \sup_{t \geq0}\left\{ \up{\alpha}(s)-\xi(s)\right\}  =\verDev{\up{\alpha}}{\xi} \numberthis \label{vdev:step5}.
	\end{align*}
	In the last line (Eq.~\eqref{vdev:step5}), we took the supremum over a larger set, so it can potentially increase. On the other hand,
	we also have that 
	\begin{align*}
		q(t) 
		\overset{\eqref{eq:backlog}}{=}& A(t)-D(t)\\ 
		{\leq} & A(t)  \numberthis \label{vdev:step2}\\
		\overset{\eqref{ineq:ac}}{\leq} &  \up{\alpha}(t)  \\
		{\leq} & \sup_{s\geq 0}\left\{ \up{\alpha}(s)\right\} 
	\end{align*}
	In the second line (Eq.~\eqref{vdev:step2}) we used the fact that $D \geq 0$. 
    Therefore, the backlog is less than the minimum of the two bounds. 
\end{proof}

So, the usual backlog bound from Thm.~\ref{thm:performancebounds} is almost recovered. Note, however, that the special case of a bounded arrival curve needs to be treated 
explicitly in the case of negative service curves, since the vertical deviation can be conservative for the case that the arrival curve never reaches $v(\up{\alpha}, \xi)$ (see also Fig.~\ref{fig:th19_cases}c). 

This observation indicates that proving the tightness of the backlog bound is more involved than in the standard case, where we achieve the vertical deviation by simply setting $A=\up{\alpha}$ ("greedy arrivals") and $D=\up{\alpha} \otimes \beta$ ("lazy server") \cite{bouillard2018deterministic}. The complication arises due to the fact that the vertical deviation is taken on when $\xi < 0$, yet for the actual departures we have, of course, $D\geq0$.
Hence, we need to find a worst-case sample path that actually provokes the backlog bound from Thm.~\ref{thm:backlog-bound}.

Next, we prove the tightness of the backlog bound. Here, we need to distinguish cases corresponding to the minimum $\verDev{\up{\alpha}}{\xi} \wedge\sup_{s\geq 0}\left\{ \up{\alpha}(s)\right\}$ in Thm.~\ref{thm:backlog-bound}.
Further, for ease of presentation in the proof, we make the assumption of the maximal arrival curve $\up{\alpha}$ being continuous $\forall t>0$. 

\begin{thm}[Tightness of the Backlog Bound]

\label{thm:backlog-bound-tightness} Let an arrival process $A$ traverse a system $\mathcal{S}.$  Further, let the arrivals be constrained by a sub-additive maximal arrival curve $\up{\alpha}\in\fn$, $\up{\alpha}(t)$ being continuous $\forall t>0$. The
system offers a service curve $\xi\in\mathcal{F}_{\leq\text{0}}^{\uparrow}$. 
Let $\tvd\coloneqq\arg\sup_{s\geq0}\left\{ \up{\alpha}(s)-\xi(s)\right\}$. 
We have to treat the following cases:

\underline{Case I} ("No plateau"): $\exists t\geq 0: \up{\alpha}(t)\geq\verDev{\up{\alpha}}{\xi}$.

That means we have an arrival curve which grows large enough such that it is possible for the backlog to attain $\verDev{\up{\alpha}}{\xi}$.

\underline{Case I-A} ("The standard case", see Fig.~\ref{fig:th19_cases}a):  
$\xi(\tvd)\geq0$. \\
Set \textup{$\awc\coloneqq\up{\alpha}$}
and \textup{$\dwc\coloneqq \posPart{\up{\alpha}\otimes\xi}$,} then,

\begin{equation}
q(\tvd)=\verDev{\up{\alpha}}{\xi} \wedge\sup_{s\geq 0}\left\{ \up{\alpha}(s)\right\}.\label{eq:backlog-bound-tightness-1}
\end{equation}

In this case, the negativity of $\xi$ essentially plays no role (as the vertical deviation is attained when $\xi\geq0$, see again Fig.~\ref{fig:th19_cases}a) and the worst-case sample path is the conventional one with greedy arrivals and lazy server.

\underline{Case I-B} ("The interesting case", see Fig.~\ref{fig:th19_cases}b): $\xi(\tvd)<0$. \\
Set $t_{B}\coloneqq\up{\alpha}^{-1}(\verDev{\up{\alpha}}{\xi})$,
\textup{\\
\[ \awc(t)\coloneqq
\begin{cases}
    \up{\alpha}(t_{B})-\up{\alpha}(t_{B}-t), & \text{if $t \leq t_{B}$,} \\
    \up{\alpha}(t_{B}), & \text{otherwise,}
  \end{cases}\]}
and \textup{$\dwc\coloneqq\left[\awc\otimes\xi\right]^{+}$}. 

Then,
\begin{equation}
q(t_{B})=\verDev{\up{\alpha}}{\xi} \wedge\sup_{s\geq 0}\left\{ \up{\alpha}(s)\right\}.\label{eq:backlog-bound-tightness-2}
\end{equation}

This is the interesting case where $\verDev{\up{\alpha}}{\xi}$ is attained at a later point in time on the worst-case sample path than for arrival and service curve, because $\xi$ is still negative at time $t_{vd}$. Here, the worst-case sample path is not just greedy arrivals and lazy server.

\underline{Case II} ("The plateau case", see Fig.~\ref{fig:th19_cases}c): \\ $\up{\alpha}(t)<\verDev{\up{\alpha}}{\xi}, \forall t\geq0$ and $\exists t_p\geq0\textup{ such that }\forall t\geq t_p:\up{\alpha}(t)=p\coloneqq\sup_{s\geq0}\up{\alpha}(s)$. 
Set \textup{\\
\[ \awc(t)\coloneqq
\begin{cases}
    \up{\alpha}(t_{p})-\up{\alpha}(t_{p}-t), & \text{if $t \leq t_{p}$,} \\
    \up{\alpha}(t_{p}), & \text{otherwise,}
  \end{cases}\]}
and \textup{$\dwc\coloneqq\left[\awc\otimes\xi\right]^{+}$}. Then,
\begin{equation}
q(t_{p})=\verDev{\up{\alpha}}{\xi} \wedge\sup_{s\geq 0}\left\{ \up{\alpha}(s)\right\}.\label{eq:backlog-bound-tightness-3}
\end{equation}

The special case of an arrival curve with a plateau needs to be dealt with separately, since the backlog may never attain $\verDev{\up{\alpha}}{\xi}$, when the plateau is not high enough.

\end{thm}
\begin{proof} 

Let us consider \underline{Case I-A}, then:

\begin{align*}
q(\tvd)  \overset{\eqref{eq:backlog}}{=}&\awc(\tvd)-\dwc(\tvd)\\
=& \up{\alpha}(\tvd)-\posPart{\up{\alpha}\otimes\xi(\tvd)}\\
=&\up{\alpha}(\tvd)-\up{\alpha}\otimes\xi(\tvd)\numberthis \label{ver_tight:step3}\\
\geq& \up{\alpha}(\tvd) - \xi(\tvd)\numberthis \label{ver_tight:step4}\\
=& \verDev{\up{\alpha}}{\xi}\\
=&\verDev{\up{\alpha}}{\xi}\wedge \sup_{s\geq0}\up{\alpha}(s).
\end{align*}

Due to $\xi(\tvd)\geq0$, the third Eq.~\eqref{ver_tight:step3} holds. In the fourth step (Eq.~\eqref{ver_tight:step4}) we used the fact that $\xi=\delta_0\otimes\xi\geq\up{\alpha}\otimes\xi$, since $\up{\alpha}(0)=0$ and the isotonicity of the convolution (see Remark~\ref{rem:conv}). Then, by the upper bound on the backlog from Thm.~\ref{thm:backlog-bound} and $\sup_{s\geq0}\up{\alpha}(s)\geq\verDev{\up{\alpha}}{\xi}$, the claim follows.

The arrival and service curve properties as well as causality are obviously fulfilled since we are in the standard case.

For \underline{Case I-B}, we first check that the sample path $\awc$ is conforming to the maximal arrival curve $\up{\alpha}$.

To that end, it suffices to verify the maximal arrival curve property in the interval $[0,t_{B}]$, since for $t>t_{B}$, $\awc(t)$ is trivially conforming. Hence,  $\forall s,t\in[0,t_{B}] \textup{ with } s\leq t$:
\begin{align*}
    \awc(t)-\awc(s)=& \up{\alpha}(t_{B})-\up{\alpha}(t_{B}-t) -\\& (\up{\alpha}(t_{B})-\up{\alpha}(t_{B}-s)) \\
    =& \up{\alpha}(t_{B}-s)-\up{\alpha}(t_{B}-t) \\
    \overset{\eqref{ineq:subadditive}}{\leq} & \up{\alpha}(t-s).
\end{align*}

Next, we show that 
\begin{equation}\label{eq:dwc_tb}
    \dwc(t_{B})=0.
\end{equation} For this, it suffices to
show that $\conv{\awc}{\xi}{t_{B}}=0$:  
\begin{align*}
\conv{\awc}{\xi}{t_{B}}= & \inf_{0\leq s\leq t_{B}}\left\{ \awc(t_{B}-s)+\xi(s)\right\} \\
= & \inf_{0\leq s\leq t_{B}}\left\{ \up{\alpha}(t_{B})-\up{\alpha}(s)+\xi(s)\right\} \\
\overset{\eqref{eq:minsupinf}}{=} & \up{\alpha}(t_{B})-\sup_{0\leq s\leq t_{B}}\left\{ \up{\alpha}(s)-\xi(s)\right\} \\
= & \verDev{\up{\alpha}}{\xi}-\verDev{\up{\alpha}}{\xi}=0.
\end{align*}

In the last line, we used the definition of $t_{B}$ and
the fact that $\tvd\leq t_{B}$ in this case, since $\xi(\tvd)<0$ and thus $\up{\alpha}(\tvd)\leq \up{\alpha}(\tvd) - \xi(\tvd)=\verDev{\up{\alpha}}{\xi}=\up{\alpha}(t_{B})$, and $\up{\alpha}$ being non-decreasing. 

Then, we obtain
\begin{align*}
    q(t_{B})\overset{\eqref{eq:backlog}}{=}&\awc(t_{B})-\dwc(t_{B})\overset{\eqref{eq:dwc_tb}}{=}\awc(t_{B})=\up{\alpha}(t_{B})\\
    =&\up{\alpha}(\up{\alpha}^{-1}(\verDev{\up{\alpha}}{\xi})\\
    =&\verDev{\up{\alpha}}{\xi}\numberthis \label{ver_tight:pseudo}\\=&\verDev{\up{\alpha}}{\xi}\wedge \sup_{s\geq0}\up{\alpha}(s),
\end{align*}

\begin{figure*}
    \centering
    \includegraphics[width=0.75\textwidth]{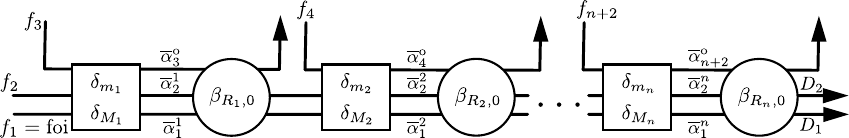}
    \caption{General form of a C/C system with three flows crossing each C/C component.}\label{fig:compcommgeneral}
\end{figure*}

where in the second to last step (Eq.~\eqref{ver_tight:pseudo}) the pseudo-inverse (Def.~\ref{def:pseudo}) is exact, due to the continuity of $\up{\alpha}$, and in the last step we use the same argument as in the last step of \underline{Case I-A}.
We note that due to Th.~\ref{thm:backlog-bound}, $\forall t\geq0$:\[ q(t)\leq \verDev{\up{\alpha}}{\xi} \wedge\sup_{s\geq 0}\left\{ \up{\alpha}(s)\right\} = q(t_{B}).\]

 $\dwc$ is clearly conforming to the service curve $\xi$. Further, we created a system which is causal, i.e. $\awc\geq \dwc\:$
since $\awc\in\mathcal{F_{\text{0}}^{\uparrow}}$ and thus $\awc\geq\posPart{\awc\otimes\xi}$ (using again the special case of the isotonicity of the convolution).

Lastly, we treat \underline{Case II}: clearly $p<\verDev{\up{\alpha}}{\xi}$. Again, $\awc$ is conforming to the maximal arrival curve $\up{\alpha}$ (due to the sub-additivity of $\up{\alpha}$, see \underline{Case I-B}).

We show that  
\begin{equation}\label{eq:dwc_tp}
    \dwc(t_{p})=0, 
\end{equation}
for which it is sufficient to show that $\conv{\awc}{\xi}{t_{p}}<0$:
\begin{align*}
\conv{\awc}{\xi}{t_{p}}= & \inf_{0\leq s\leq t_{p}}\left\{ \awc(t_{p}-s)+\xi(s)\right\} \\
= & \inf_{0\leq s\leq t_{p}}\left\{ \up{\alpha}(t_{p})-\up{\alpha}(s)+\xi(s)\right\} \\
\overset{\eqref{eq:minsupinf}}{=} & \up{\alpha}(t_{p})-\sup_{0\leq s\leq t_{p}}\left\{ \up{\alpha}(s)-\xi(s)\right\} \numberthis \label{ver_tight:case2}.
\end{align*}

 To continue with Eq.~\eqref{ver_tight:case2}, we need to distinguish two cases:  (a) if $t_{p}\geq \tvd$, we have
\begin{align*}
\up{\alpha}(t_{p})-\sup_{0\leq s\leq t_{p}}\left\{ \up{\alpha}(s)-\xi(s)\right\} 
= p - \verDev{\up{\alpha}}{\xi} < 0;
\end{align*}
(b) if $t_{p}<\tvd$, we have $\xi(t_{p}) \leq \xi(t_{vd})$ (as $\xi$ is non-decreasing) and $\alpha(t_{p}) = \alpha(t_{vd})=p$. This implies 
\begin{align*}p - \xi(t_p) &= \alpha(t_{p})-\xi(t_{p}) \\
& \geq \alpha(t_{vd})-\xi(t_{vd})=\verDev{\up{\alpha}}{\xi}.
\end{align*}
For $p<\verDev{\up{\alpha}}{\xi}$, we see 
$\xi(t_{p})<0,$ and thus
$\forall t\leq t_{p}, \xi(t)<0$ (as $\xi$ is non-decreasing). We continue with Eq.~\eqref{ver_tight:case2}:
\[\up{\alpha}(t_{p})-\sup_{0\leq s\leq t_{p}}\left\{ \up{\alpha}(s)-\xi(s)\right\} < \up{\alpha}(t_{p})-\sup_{0\leq s\leq t_{p}}\left\{ \up{\alpha}(s)\right\}=0. 
\]

Thus, we obtain
\begin{align*}
    q(t_{p})\overset{\eqref{eq:backlog}}{=}&\awc(t_{p})-\dwc(t_{p})\overset{\eqref{eq:dwc_tp}}{=}\awc(t_{p})\\
    =&\up{\alpha}(t_{p})=p=\verDev{\up{\alpha}}{\xi}\wedge p\\=&\verDev{\up{\alpha}}{\xi} \wedge\sup_{s\geq 0}\left\{ \up{\alpha}(s)\right\},
\end{align*}
 where in the second to last step we use $p<\verDev{\up{\alpha}}{\xi}$.
 We note that due to Th.~\ref{thm:backlog-bound}, $\forall t\geq0$:\[ q(t)\leq \verDev{\up{\alpha}}{\xi} \wedge\sup_{s\geq 0}\left\{ \up{\alpha}(s)\right\} = q(t_{p}).\]
 
$\dwc$ is clearly conforming to the service curve $\xi$.
Further, we created a system which is causal, i.e. $\awc\geq \dwc$
since $\awc\in\mathcal{F_{\text{0}}^{\uparrow}}$ and thus $\awc\geq\posPart{\awc\otimes\xi}$ (again by the special case of the isotonicity of the convolution).
\end{proof}

\section{Patterns of Application}\label{sec:apps}

With a broader set of service curves that we can derive performance bounds from, the question of potential applications arises. The extended NC results remove a previous blind spot, where a system with multiple flows but no \emph{strict} aggregate service curve could not be adequately modeled and analyzed. We are now also able to exploit the concatenation theorem (see Thm.~\ref{def:concat}) while still obtaining performance bounds in a system that would normally rely on a strict service curve. This is desirable, as a node-by-node analysis often cannot capture certain properties of the overall system \cite{schmitt2008delay},\cite[Section 1.4.3]{le2001network}, resulting in less accurate performance bounds.

In this section, we examine two possible patterns of applications where the novel results provide an interesting insight into the network performance analysis. We show that the MinAC analysis can improve on results of state-of-the-art techniques and may even enable a network analysis for certain areas of the parameter space where existing techniques deliver no solution at all. 

\subsection{Computation-Communication Systems}\label{sub:app1}

We consider a pattern of a mixed Computation-Communication (C/C) system consisting of $n$ components and $n+2$ flows (see Fig.~\ref{fig:compcommgeneral}), and proceed with deriving formulas for calculating the end-to-end delay bound across the $n$ components. Let $f_1$ be the flow of interest and $f_2$ a cross-flow. Both flows traverse components $1,\dots, n$ as an aggregate. Assume that there are $n$ additional cross-flows $f_i,i\in\{3,n+2\}$, passing through each C/C component $i-2$, respectively. Each flow is constrained by a maximal token-bucket arrival curve $\up{\alpha}_i=\gamma_{r_i,b_i}$. The foi is additionally restricted by a minimal arrival curve $\mac{\alpha}_1=\rlac$. We assume that the delay at each computational element $i\in\{1,n\}$ in the system is lower bounded by $m_i$ and upper bounded by $M_i$ (modeled in Fig.~\ref{fig:compcommgeneral} as service curves using the shift function $\delta$ as in \cite[Theorem 6.2]{bouillard2018deterministic}). Let $T_i:=M_i-m_i$ be the delay variance at each computational element $i$. Each communication element $i\in\{1,n\}$ provides a simple constant-rate service curve $\beta_{R_i,0}$. We assume static priority scheduling at each communication element, and assign the highest priority to flows $f_i,i\in\{3,n+2\}$. Flow $f_2$ is assigned the second highest priority, and the foi the lowest. Using the results proposed in Sect.~\ref{sec:bounds} for our MinAC analysis (mac), we first calculate the residual service curve for flow $f_1$ as 
\begin{align*}
    \ma\coloneqq& \left(\left(\bigotimes_{i=1}^n \left(\beta_{R_i,T_i}-\up{\alpha}_{i+2}\right) \right) - \up{\alpha}_2 \right)_\downarrow \\
    =& \xi_{b_2+b_{i+2}+(r_2+r_{i+2})\sum_{i=1}^n T_i, \bigwedge_{i=1}^n (R_i-r_{i+2}) - r_2, \sum_{i=1}^n T_i},
\end{align*}
with 
\begin{equation*}
    \xi_{b_N, R, T}(t)\coloneqq \rl(t)-b_N.
\end{equation*}

An end-to-end delay bound for flow $f_1$ using the MinAC analysis is then calculated as
\begin{equation}\label{eq:e2enew}
    \madelay=h(\up{\alpha}_1,\ma)\vee z(\mac{\alpha}_1, \ma).
\end{equation}

For the conventional analysis (ca), we determine the input to the C/C components by using the output bound (see Eq.~\eqref{eq:outputbound}), as the input to each component is the output of the respective flow at the preceding component. For the communication element of component $i$, the input flows are thus given as
\begin{equation*}
    \hspace*{-42mm} \up{\alpha}_2^1=\up{\alpha}_2\oslash\delta_{T_1}, \up{\alpha}_1^1=\up{\alpha}_1\oslash\delta_{T_1}, 
\end{equation*}
\begin{equation*}
    \hspace*{-4mm} \up{\alpha}_2^i=\up{\alpha}_2^{i-1}\oslash[\beta_{R_{i-1},0}-(\up{\alpha}_{i+1}\oslash \delta_{T_{i-1}})]^+ \oslash \delta_{T_i} = \gamma_{r_2^i, b_2^i},
\end{equation*}
\begin{equation*}
    \hspace*{-3mm} \up{\alpha}_1^i= \up{\alpha}_1^{i-1}\oslash[\beta_{R_{i-1},0}-((\up{\alpha}_2^{i-1}+\up{\alpha}_{i+1})\oslash\delta_{T_{i-1}})]^+ \oslash \delta_{T_i},
\end{equation*}
\begin{equation*}
\hspace*{-32mm} \up{\alpha}_{i+2}^o=\gamma_{r_{i+2},b_{i+2} + r_{i+2}\cdot T_i}=\gamma_{r_{i+2}^o, b_{i+2}^o}.
\end{equation*}
Next, the residual service curve for flow $f_1$ employing a static priority policy can be calculated for each component $i$ as
\begin{equation*}
    \cai\coloneqq \left[\beta_{R_i,0} - \up{\alpha}_2^i - \up{\alpha}_{i+2}^o \right]^+ = \beta_{\rat, \cat}.
\end{equation*}
A delay bound for communication component $i$ is obtained by 
\begin{equation*}
    h(\up{\alpha}_1^i, \cai)=\frac
    {b_1+r_1\sum_{j=1}^i T_j + r_1\sum_{j=1}^{i-1}\frac{b_2^j + b_{j+2}^o}{R_{\mathrm{res}}^{\mathrm{ca},j}}}{\rat} + \cat 
\end{equation*}
A delay bound for the whole C/C component $i$ is simply   $d_i=T_i + h(\up{\alpha}_1, \cai)$. An end-to-end delay bound for flow $f_1$ using the conventional analysis is then calculated as
\begin{equation}\label{eq:e2enodal}
    \cadelay=\sum_{i=1}^n d_i.
\end{equation}
\begin{figure}[b]
    \centering
    \includegraphics[width=0.6\columnwidth]{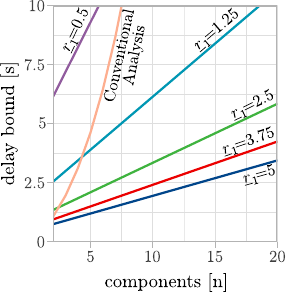}
    \caption{Comparison of delay bounds for varying rates of $\mac{r}_1$.}
    \label{fig:mixedresourcesims}
\end{figure}

\setcounter{figure}{8}

\begin{figure*}
     \centering
     \includegraphics[width=\textwidth]{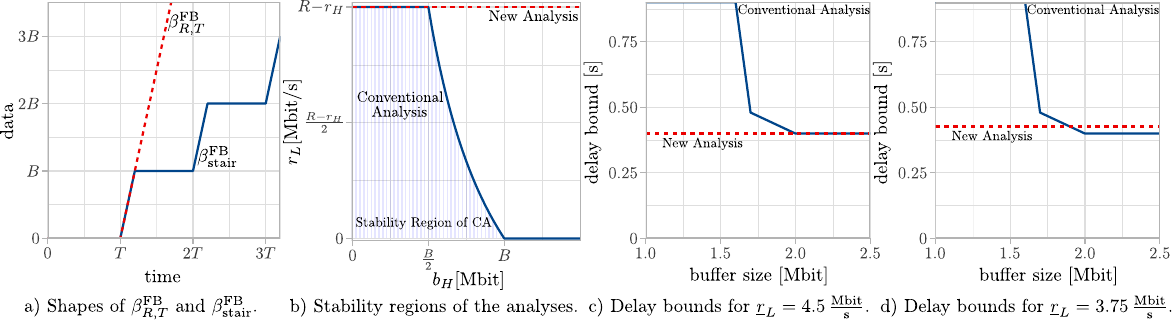}
     \caption{Illustration of finite buffer service curves, stability regions and delay bounds for the case study.}
     \label{fig:finitebufferevals}
\end{figure*}

Equipped with these formulas, we proceed with a small case study, evaluating the two analyses. To this end, we consider the general system previously described (see Fig.~\ref{fig:compcommgeneral}). We calculate the end-to-end delay bound for the foi $f_1$ in this system. To evaluate the effect of the minimal arrival curve, we define a general parameter set and vary the minimal rate $\mac{r}_1$ over a range of values. Let $b_1=b_2=b_3=\mb{1}$, $r_1=r_2=r_3=\mbps{5}$, $R_i=\mbps{20}\eqqcolon R$, and $T_i=\ms{50}$, $i \in \{1,\dots,n\}$. We set $T_{\mac{\alpha_1}}=\frac{b_1}{R}$ and choose $\mac{r}_1\in\left\{0.5, 1.25, 2.5, 3.75, 5\right\}\mbps{}$. The delay bound is calculated for different numbers of C/C components $n\in\{2\dots, 20\}$.
For each value of $\mac{r}_1$ and $n$, we calculate the end-to-end delay bounds for the MinAC analysis using Eq.~\eqref{eq:e2enew}, and for the conventional analysis using Eq.~\eqref{eq:e2enodal}. The results are shown in Fig.~\ref{fig:mixedresourcesims}. We can see that for $\mac{r}_1\in\{\mbps{3.75},\mbps{5}\}$, we always achieve a much more accurate end-to-end delay bound. For $\mac{r}_1\in\{\mbps{1.25},\mbps{2.5}\}$, the MinAC delay bound is below the conventional delay bound from $5$ resp. $3$ C/C components in the system onwards. For \mbps{0.5}, however, the new end-to-end delay bound becomes more conservative for small numbers of  components. Expectedly, with our newly proposed approach, we rely on the guarantees provided by a minimal arrival curve. Consequently, the better the guarantees, i.e., the higher $\mac{r}_1$, the better the calculated end-to-end delay bound becomes. 
However, we can observe in Fig.~\ref{fig:mixedresourcesims} that even for the smallest minimum arrival rate of \mbps{0.5} we have a better scaling of the delay bound than for the conventional analysis, which exhibits a super-linear scaling in the number of components.  
This means that, for large enough systems, the MinAC approach will eventually outperform the conventional analysis, even with low minimal arrival guarantees. 

\subsection{Finite Shared Buffers}\label{sub:app2}
\setcounter{figure}{7}
\begin{figure}[b]
    \centering
    \includegraphics[width=0.8\columnwidth]{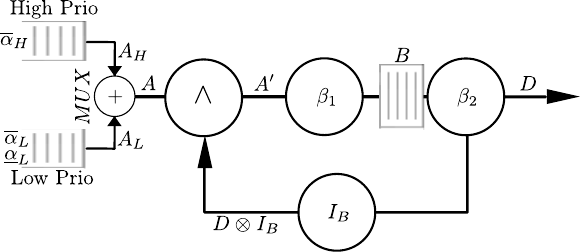}
    \caption{System with a finite shared buffer.}
    \label{fig:finitebuffer}
\end{figure}

In this application pattern, we consider two priority queues, one for a high and the other for a low priority flow. First, we derive the required buffer sizes for each flow and find that conventional NC analyses cannot properly express and analyze all feasible system designs. Next, we calculate delay bounds for the low priority foi $f_L$ in this finite shared buffer system.

Before we can derive performance bounds in the system for the low priority flow $f_L$, we need to determine the residual service curve for both the conventional and MinAC analysis. Consider the system in Fig.~\ref{fig:finitebuffer}. Let $I_B$ be the service curve of the feedback control for arrivals exceeding the finite buffer with capacity $B$ at $\beta_2$. We define $I_B(t)=+\infty$ for $t>0$ and $I_B(0)=B$, as in \cite{agrawal1999performance, bouillard2018deterministic}, \cite[Section 2.3.7]{chang2000performance}. It holds that $D \otimes I_B(t)=D(t)+B$, and, hence, $A'$ cannot be more than $B$ data units ahead of $D$ (as $A'=A\wedge (D \otimes I_B)$). We let $\beta_i=\beta_{R_i,0}, i=1,2$, $R=R_1\wedge R_2$, and $T=T_1+T_2$. For this closed-loop feedback system, we have 
\begin{equation*}
A' \geq A \wedge (D\otimes I_B), D\geq A'\otimes\beta_1\otimes\beta_2,
\end{equation*}
where $A=A_H+A_L$. Combining both inequalities, we obtain 
\begin{equation*}
    D\geq A\otimes(\beta_1\otimes\beta_2) \wedge D\otimes(I_B\otimes\beta_1\otimes\beta_2),
\end{equation*}
which can be turned into an open-loop system \cite[Section 2.3.7]{chang2000performance}  
\[D\geq A\otimes(\beta_1\otimes\beta_2) \otimes (I_B\otimes\beta_1\otimes\beta_2)^*,\]
where $(I_B\otimes\beta_1\otimes\beta_2)^*$ is the sub-additive closure (see Def.~\ref{def:subadd-closure}). Hence, the system offers a service curve $\beta^{\text{FB}}=\beta_1\otimes\beta_2\otimes(I_B\otimes\beta_1\otimes\beta_2)^*$. In general, it holds that, for $RT\leq B$, the service curve offered to the flow is equal to $\beta^{\text{FB}}_{R,T}=\beta_1\otimes\beta_2$. If, however, the bandwidth-delay product $RT$ is greater than the available buffer $B$, it holds that the service curve is a staircase function $\beta^{\text{FB}}_{\text{stair}}$, since there is not enough buffer space available to serve the flow without delaying it at the entrance to the feedback loop. Both service curves are illustrated in Fig.~\ref{fig:finitebufferevals}a. 

In the following, we assume that both flows are upper-constrained by token buckets $\up{\alpha}_H = \gamma_{r_H,b_H}$ and $\up{\alpha}_L = \gamma_{r_L,b_L}$, respectively. We let $T_i=0$ for $\beta_1$ and $\beta_2$. Consequently, it always holds that $\beta^{\text{FB}}=\beta^{\text{FB}}_{R,T}=\beta_{R,0}$ for the aggregate of the flows. Thus, the buffer requirement for the high priority queue is equal to
\begin{equation*}
    v(\up{\alpha}_H, \beta^{\text{FB}})=v(\gamma_{r_H,b_H}, \beta_{R,0}) = b_H,
\end{equation*}
independent of which analysis we choose. This is not the case for flow $f_L$, though. For this, we first have to calculate the residual service curves for each analysis. For the MinAC analysis, the residual service curve is calculated as
\begin{align*}
    \ma=& \left(\beta^{\text{FB}}-\up{\alpha}_H\right)_\downarrow = \left(\beta_{R,0} - \up{\alpha}_H \right)_\downarrow = \xi_{b_H, R- r_H, 0}.
\end{align*}
Note that $\ma$ is independent of the feedback control, i.e., its shape does not depend on the relation of $RT$ and $B$. In contrast, the residual service curve for the conventional analysis does depend on this relation. We calculate it as
\begin{align*}
    \ca=&\ \left[\beta_1\otimes\beta_2-\up{\alpha}_H\right]^+ \otimes (\left[\beta_1\otimes\beta_2-\up{\alpha}_H\right]^+\otimes I_{B-v(\up{\alpha}_H,\beta)})^* \\
    =&\ \beta_{R-r_H, \frac{b_H}{R-r_H}}\otimes (\beta_{R-r_H, \frac{b_H}{R-r_H}}\otimes I_{B-b_H})^*.
\end{align*}
Using each residual service curve, we determine the buffer requirement for the low priority queue. For the MinAC analysis, we obtain
\begin{equation}\label{eq:nabuffer}
    v(\up{\alpha}_L, \ma)=b_H+b_L.
\end{equation}

For the conventional analysis, we need to consider the relation of the bandwidth-delay product for the residual feedback system $\rlo\tlo=(R-r_H)\frac{b_H}{R-r_H}=b_H$, and the buffer available in it for the low priority flow $\blo=B-b_H$.
If $\rlo\tlo\leq \blo$, i.e., $b_H \leq B-b_H$, then the residual service curve $\ca$ follows the shape of $\beta^{\text{FB}}_{R,T}$ (see Fig.~\ref{fig:finitebufferevals}a). 
In this case, i.e., $b_H\leq\frac{B}{2}$, we obtain that 
\begin{equation*}
    v(\up{\alpha}_L, \ca) = v(\gamma_{r_L,b_L}, \beta_{R-r_H, \frac{b_H}{R-r_H}}) =
    b_L+r_L\frac{b_H}{R-r_H}.
\end{equation*}
For $\rlo\tlo > \blo$, i.e. $b_H>\frac{B}{2}$, we follow the shape of $\beta^{\text{FB}}_{\text{stair}}$, and discover an interesting restriction regarding the rate $r_L$ of flow $f_L$ in order to not diverge from $\ca$:
\begin{equation}\label{eq:diverge}
    r_L\leq \frac{\blo}{\tlo} = \left(\frac{B}{b_H}-1\right)(R-r_H).
\end{equation}
We give a brief intuition for Eq.~\eqref{eq:diverge}. $\frac{\blo}{\tlo}$ is the long-term rate of $\ca$. We can only calculate finite bounds if $\alpha_L$ and $\ca$ do not diverge. If $r_L>\frac{\blo}{\tlo}$, i.e., the rate of $\alpha_L$ is larger than the long-term rate of $\ca$, then the stability of the system is not ensured and infinite performance bounds result.

As $\blo= B - b_H$, we recognize that for $b_H\geq B$, we cannot compute a backlog bound for $f_L$ using the conventional residual service curve $\ca$. Furthermore, as we see in Eq.~\eqref{eq:diverge}, the feasible rate $r_L$ decreases hyperbolically in $b_H$ over the interval $\left(\frac{B}{2},B\right)$, further limiting the ability to calculate a backlog bound. In Fig.~\ref{fig:finitebufferevals}b, the so-called stability region of the conventional analysis is shown. Here, the stability region is the parameter space for which we can compute finite backlog bounds.

Clearly, the closer the buffer is to being full with traffic of $f_H$, the less $f_L$ can send in each window interval $[(i-1)\tlo,i\tlo), i>1$, eventually diverging from $\ca$. As a result, we cannot determine the vertical deviation for arbitrary $r_L$ that would be valid under the "normal" stability condition $r_L\leq R-r_H$, but violate Eq.~\eqref{eq:diverge}. In conclusion, the conventional analysis is not able to provide a backlog bound for arbitrary flows $f_H, f_L$. 
In contrast, the calculation of the buffer requirement based on the MinAC analysis in Eq.~\eqref{eq:nabuffer} is only restricted by $r_L\leq R-r_H$, thus resulting in a much larger stability region (see again Fig.~\ref{fig:finitebufferevals}b). 

We move on to the delay bound calculation. For flow $f_H$, the delay bound calculation is the same for both analyses:
\begin{equation*}
    h(\up{\alpha}_H, \beta^{\text{FB}})
    =h(\gamma_{r_H,b_H}, \beta_{R,0})
    =\frac{b_H}{R}.
\end{equation*}
For flow $f_L$, this looks different, as we have different residual service curves. For the new anaysis, assuming $\mac{\alpha}_L = \beta_{r_L, T_{\mac{\alpha}_L}}$, we calculate
\begin{align*}\label{eq:fbnewdelay}
    \madelay & = h(\up{\alpha}_L,\xi_{b_H, R-r_H,0}) \vee z(\mac{\alpha}_L, \xi_{b_H, R-r_H,0}) \\
    & = \left(\frac{b_H+b_L}{R-r_H} \right) \vee \left(T_{\mac{\alpha}_L} + \frac{b_H}{\mac{r}_L} \right). \numberthis
\end{align*}
For the conventional analysis, if we have a staircase residual service curve, we calculate the number of stairs that are needed in the delay bound calculation as $i^* \coloneqq \lceil\frac{ b_L }{B-b_H}\rceil$, and obtain
\begin{equation}\label{eq:fbconvdelay}
    \cadelay=\begin{cases}
      \frac{b_H+b_L}{R-r_H}, \hspace*{-2mm}& \rlo\tlo\leq \blo,\\
      \frac{b_L-\left(i^*-1\right)(B-b_H)}{R-r_H} + i^*\tlo, \hspace*{-2mm}& \text{otherwise},
    \end{cases}
\end{equation}
where Eq.~\eqref{eq:diverge} and $b_H<B$ have to hold (see also Fig.~\ref{fig:finitebufferevals}b again), otherwise, $h(\up{\alpha}_L, \ca)=\infty$. 

We proceed with a brief case study on the two approaches to calculating delay bounds. Consider again the system in Fig.~\ref{fig:finitebuffer}. Let $b_L=\mb{2}, b_H=\mb{1},$ and $r_L=r_H=\mbps{5}$. For $\mac{\alpha}_L$, we let $T_{\mac{\alpha}_L}=\frac{b_L}{R}$ and $\mac{r}_L\in\{\mbps{3.75}, \mbps{4.5}\}$. Each system offers a service curve $\beta_i=\beta_{R_i,0}$ with $R_i=\mbps{12.5}$. We calculate the delay bound using these parameter values for both approaches, varying the size of the finite buffer $B$. The results are given in Fig.~\ref{fig:finitebufferevals}. For $\mac{r}_L=\mbps{4.5}$ (Fig.~\ref{fig:finitebufferevals}c), we see that the delay bound of the MinAC analysis is always either equal to or more accurate than the conventional analysis. For $\mac{r}_L=\mbps{3.75}$ (Fig.~\ref{fig:finitebufferevals}d), the delay bound calculation using Eq.~\eqref{eq:fbnewdelay} falls into the second case of the maximum operator. As a result, for $b_H\leq \frac{B}{2}$, the conventional analysis achieves slightly more accurate bounds. However, for $b_H>\frac{B}{2}$, this changes, as the conventional analysis now calculates its delay bound using the second case of Eq.~\eqref{eq:fbconvdelay}, instead. Now, the delay bound becomes much larger than for the MinAC analysis.

\section{Conclusion}

In this report, we extended the NC framework to deal with scenarios in which an aggregate min-plus service curve is given and we want to calculate residual service curves in order to compute per-flow performance bounds. In this case, partially negative and decreasing service curves arise and existing NC results on performance bounds cannot be applied. We remove this blind spot with the aid of minimal arrival curves, which allow us to calculate tight performance bounds for negative service curves.

This generalization of the performance bounds for negative service curves leads to more flexibility in the modeling of applications, while requiring different assumptions about the system. We effectively trade the strictness of the service curve for the existence of a minimal arrival curve. 
Using the new NC results, we have shown that we can improve the performance analysis of interesting application patterns; we are even able to analyze systems for which a conventional analysis fails to provide performance bounds.

\balance

\bibliographystyle{IEEEtran}	
\bibliography{refs2.bib}
\end{document}